# Dual-Use Commercial and Military Communications on a Single Platform using RAN Domain Specific Language

Alan Gatherer
Cirrus360
Dallas, USA
gatherer@cirrus3sixty.com

Chaitali Sengupta
Cirrus360
Dallas, USA
chaitali@cirrus3sixty.com

Sudipta Sen
Cirrus360
Dallas, USA
sudipta@cirrus3sixty.com

Jeffery H. Reed
Cirrus360 and Virginia Tech
Blacksburg, USA
reedjh@cirrus3sixty.com

*Abstract*— **Despite the success of the O-RAN Alliance in developing a set of interoperable interfaces, development of unique Radio Access Network (RAN) deployments remains challenging. This is especially true for military communications, where deployments are highly specialized with limited volume. The construction and maintenance of the RAN, which is a real time embedded system, is an ill-defined NP problem requiring teams of specialized system engineers, with specialized knowledge of the hardware platform. In this paper, we introduce a RAN Domain Specific Language (RDSL™) to formally describe use cases, constraints, and multi-vendor hardware/software abstraction to allow automation of RAN construction. In this DSL, system requirements are declarative, and performance constraints are guaranteed by construction using an automated system solver. Using our RAN system solver platform, Gabriel™ we show how a system engineer can confidently modify RAN functionality without knowledge of the underlying hardware. We show benefits for specific system requirements when compared to the manually optimized, default configuration of the Intel FlexRAN™, and conclude that DSL/automation driven construction of the RAN can lead to significant power and latency benefits when the deployment constraints are tuned for a specific case. We give examples of how constraints and requirements can be formatted in a "Kubernetes style" YAML format which allows the use of other tools, such as Ansible, to integrate the generation of these requirements into higher level automation flows such as Service Management and Orchestration (SMO).**

## I. Introduction

The Department of Defense (DoD) intends to adopt 5G to the greatest extent possible. Nevertheless, 5G was developed as a commercial system, and adapting it to diverse and distinct military applications requires flexibility throughout the network. Military and first responder deployments of 5G have unique spectral use requirements, and add functionality that is not part of the commercial 5G/6G standard to make the system more secure, or to add sensing and radar capabilities to the network. We present a way to accomplish the flexibility needed by the military by using domain-specific languages combined with O-RAN to create a development environment that facilitates easy customization of the Radio Access Network (RAN). Our approach builds on the open RAN ecosystem created by the O-RAN Alliance for building an open and disaggregated network.

The open RAN ecosystem and the O-RAN Alliance has made tremendous strides toward a more open and disaggregated Radio Access Network (RAN) through the introduction and detailed definition of interfaces that allow multiple vendors' products to communicate in a plug-and-play manner. Disaggregation provides opportunities for new and existing vendors to innovate and provide cost effective solutions for a growing set of new use cases and markets.

However, the journey toward disaggregation and automation is far from complete. This has become increasingly clear as open RAN deployments has progressed in the last two years [1]. Open RAN systems have allowed the disaggregation of the RAN into a limited number of multi-vendor Radio, Distributed and Centralized Unit (RU, DU and CU) black boxes, and has allowed the use of Apps in the RAN Intelligent Controller (RIC) [3][4] from a wider set of vendors. But, the DU, CU and RU remain as black boxes; components in the system that are generally left alone. Of course, someone must develop these black boxes. Outside of the few large Original Equipment Manufacturers (OEMs) proprietary solutions, the O-RAN Distributed Unit (DU) has a few canonical implementations with the Open Air Interface (OAI) [16] and Software Radio Systems (SRS) [17] solutions as well as the Intel FlexRAN™ platform [10] and more recently the Nvidia Aerial platform. But generally these DU are used with little to no modification, and there has been no explosion of DU solutions on the market, or an explosion of new hardware platforms to support the existing open source code. Systems Integration (SI) teams who inherit these code bases, struggle to make even minor changes without errors appearing that require sophisticated debug. As a result these SI have invested in teams of experts with specialization in the hardware platforms, the middleware and software whose job it is to maintain the stability of the code as requirements change. This made it hard for small companies to innovate in this area and the result is that there remain very few sources for production DU. For the Radio Unit (RU) the picture is even bleaker [5]. For the Centralized Unit (CU) and 5G Core there is more variety to choose from.

The lack of variety in DU and RU solutions, and the difficulty in modifying these to meet new requirements, is an especially concerning problem for the military communications

This material is based upon work supported in part by an NTIA PWSCIF award under award number 48-60-IF006

community. They are committed to reuse of commercial technology, 5G in particular, but must also add new functionality that is of a sensitive nature. Some examples are the addition of new security features by modification of the 5G physical layer algorithms [12], QoS enhancements for in theater robotic surgery [11], ultra-reliable low-latency communications for mission critical data [13] and the addition of sensing and radar to 5G [14].

So why are the DU and RU so hard to innovate? The answer lies in the intrinsic difficulty of constructing a real time system from its components, on a memory and processing limited platform, while meeting all of the constraints from the system, including timing and power. This is a difficult and specialized expertise that requires "tribal knowledge" of the RAN functionality and the hardware platform being used.

The problem is significantly worse in an operations environment because the RANs must run at scale for extended periods of time. That is to say they must be "High Availability" (HA) embedded systems. For example The Dallas Fort Worth area is about 24000 km$^2$. If we assume a cell tower every 5 km$^2$ on average, with the equipment supporting each cell tower having a 10 year lifespan, then one failure in the lifetime of the equipment will lead to about 9 truck rolls per week. Hence the DU/RU, and therefore their software, must be specified to have less than a single failure in the lifetime of the equipment hardware. Many failures are hard to find during lab testing because of the enormous state space of the DU operation, which spans every possible user data combination along with all of the signaling and measurement computations. Failures may be caused by cache misses or buffer overflows that were not tested for in the lab and appear in the field, often as Heisenbugs [6]. Functional failures can occur due to reads and writes of data too early or too late, or buffer overflow. Many of these failures are the result of chains of events that are hard to trace and even the best teams let errors through to the field that may appear sporadically and mysteriously during operation. Construction of a new DU or RU will usually take a large team of experts over a year to complete. Maintenance and upgrades will also consume large teams, carefully analyzing the impact of any changes to the RAN for potential failures of real time performance, due to thrashing of caches or opportunistic reordering of tasks.

So the problem set we must address is

1. To automate the construction of the RAN (to reduce the time it takes a large team of experts to properly construct a RAN by hand)

2. in a way that removes Heisenbugs by construction (to simplify and reduce testing time)

3. and can be deployment specific and upgradable with new features (to address the explosion of use cases in 5G, especially for military applications).

In this paper we present RDSL™ a Domain Specific Language (DSL) we designed specifically for RAN development. We have used RDSL™ along with our solver platform, Gabriel™ to automate 5G RAN deployment.

In section II we will outline the goals of a DSL given the problem described above. In section III we will describe the syntax of the DSL we have developed and give some examples of code and its intent. And in section IV we will give an example of the use of this DSL using a platform we have developed to use the DSL to implement optimized, deployment specific RAN solutions, to show the improvement in performance that is possible using a DSL for RAN construction.

In keeping with the goals of the O-RAN Alliance, the challenges we address in this paper are not related to radio performance such as cell capacity and massive MIMO algorithms. We focus on system implementation challenges of the RAN DU, as well as the RU and CU, such as hardware software disaggregation, power and cost, upgradability, software release maintenance and so on.

## II. THE GOALS AND BENEFITS OF A DOMAIN SPECIFIC LANGUAGE FOR RAN CONSTRUCTION

### A. Declarative description of the flow of data for the system requirements of the RAN

A Declarative language is one in which the programmer described what must be done and not how to do it. The opposite of this approach is an imperative language. Some popular examples include SQL, XML and YAML [18]. We wish to develop a description of the RAN functionality that is completely separate from the description of the hardware target. This will allow rapid porting as new and better platforms emerge to support O-RAN functionality. It also allows new functionality to be added, or two distinct applications to be merged onto a single hardware target in a way that is difficult to do for a real time system when imperative language is used. All this enables a vibrant open source community to contribute to the O-RAN stack because their contributions can be integrated and optimized via automation.

It is common for DSLs to be declarative because they are simpler and more application focused than general purpose languages, and it is therefore easier to bring the benefits of declarative language to them. Some example of declarative DSL include P4 for Software Defined Networking , TensorFlow for Machine Learning (ML). For computer system configuration in data centers there are multiple declarative languages used including Docker, Kubernetes, Ansible and Puppet. So it is safe to say that declarative languages are the wave of the future for safe, open, and reliable software at scale [7] and for DevOps code development.

### B. An Immutable Language to expose optimization opportunities for deployment specific construction and Testing

An Immutable language is one in which there are no variables, only defined or undefined data elements. The lack of change in data value exposes parallelism opportunities in the code because the optimizer does not have to consider which version of a data element it is using. In our case immutability also exposes timing optimization opportunities as we will demonstrate. Exposure of the timing also allows for automation of the testing of the RAN as well as formal correct-by-construction techniques to be used.

Haskell is the most famous immutable language in use today and Erlang is an immutable language used in telecoms. For DSLs immutability is an easy choice as the language is simple and focused and hence the lack of intuitiveness in general purpose languages such as Haskell are not present in a DSL [8].

### C. A DSL must integrate well with other languages and Automation systems, including AI/ML

The construction of a RAN is part of a larger DevOps loop that the O-RAN alliance is working to open and automate. Therefore it is important that our language can integrate with other languages and platforms in different part of the DevOps stack. In particular it must be able to take requirements from the network performance automation platform. The DSL must therefore be both machine readable and machine writable.

One important impact of this machine readability is that we can apply Large Language Models (LLMs) to facilitate informal human interaction throughout the DevOps process. This is because the RDSL™ grounds the syntax of the RAN definition allowing enforcement of LLM output for functionally complete and hallucination free control of the RAN requirements. The RDSL™ code becomes an effective digital twin of the RAN as it models a complete deployment.

### III. THE SYNTAX OF A RAN DOMAIN SPECIFIC LANGUAGE

We first define a top level logical partition of the RDSL™ syntax, formalizing the natural RAN system engineering development partition. This is shown conceptually in Fig. 1.

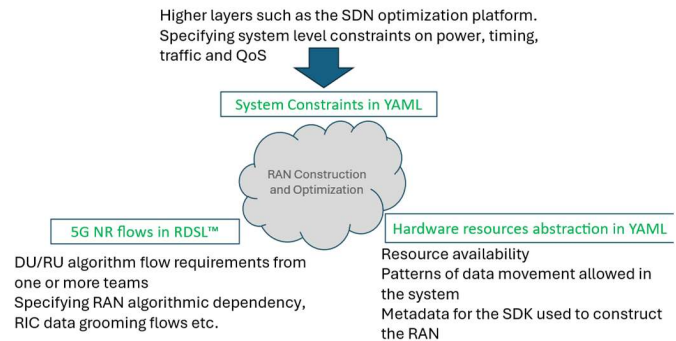

Fig. 1. Top Level Partitioning of the RAN DSL

The higher layer platform that is managing the network will provide constraints and requirements to the RAN construction platform so we must provide a syntax for this. The algorithmic definition of the flows in the RAN can be described independently of each other with the right language. This language connects tasks to each other with flow dependencies that implicitly understand the real time requirements of the RAN, and describes data management requirements in an immutable fashion (see section III.B). The language for flows requires a new language syntax which we call RDSL™. The hardware target platform is described in a YAML database format that has a predefined schema that formally defines allowed flows of data in the hardware as well as defining the resources available and their capabilities. The hardware target platform description also includes one or more packages of metadata to describe the performance of the SDK functions used to implement the flows. The performance of interfaces is similarly included in one or more packages so that the latency of data movement can be modelled. Today we include this latency as simple python code functions to maximize the flexibility of its description. This performance data is commonly parameterized by other metadata that is grounded in other parts of the description, such as by system constraints on user data rate or number of antennas and so on. But this grounding does not have to occur until the RAN construction is started and hence the hardware description can remain separate from the system definition. We will now describe each of these three syntaxes in more detail

### A. System Constraints

System constraints are defined individually in a YAML format that closely resembles Kubernetes format [9]. We mimic Kubernetes so that we can reuse as much of Kubernetes syntax management tools as possible, such as Ansible, and because RAN DevOps uses Kubernetes extensively and it is well understood in the O-RAN ecosystem. An example of two system constraints is shown in Fig. 2.

```
53  apiVersion: rdsl/v0          53  apiVersion: rdsl/v0
54  kind: timing equality        54  kind: timing equality
55  metadata:                    55  metadata:
56    name: Modem_Period2        56    name: Modem_Period
57  spec:                        57  spec:
58    equation: C <= A*370 + B < 500   58    constraint: equal
59    C: grid_period             59    variable_name: modem_period
60    A: num_ue1                 60    unit: clock
61    B: gp_base                 61    value: 1000000
62    unit: clock                62
```

Fig. 2. System Constraint Syntax Example

Each constraint is defined individually, grouped into files using the "---" separator. The apiVersion and kind fields allow the parser to interpret the specification correctly. At the time of writing there are two basic kind formats, one which creates a simple value relationship to a constant (the left example in the figure) and one which creates a more complex relationship using an equation and multiple values. Values are symbolic and can be inherited from a network/system YAML file or can be located within a flow using a label (see section III.B). In this way we can take constraints from the network and apply them hierarchically to the design of the RAN. For examples of the flexibility in applying values the reader is encouraged to look at Ansible syntax. In Fig. 2 *grid_period* is the name of a timing value. But it may be the name of a label of a timing event or stream, or may itself be defined in terms of other labels inherited from elsewhere.

### B. RDSL™

The RDSL™ flow syntax is the heart of the methodology. Flows are declarative and can be defined separately and then combined at the system level making them easy to declare and test. A flow is a data centered syntax using an immutable, infinite list of buffers, called a **stream**, as its only data type. A stream has an implicit sense of time associated with it, that allows implicit alignment of each buffer to a specific period of time. The user defines streams with the **modifiers** that create them and the streams that need to be defined in that period for the creation to occur. The **modifier** syntax defines the fine details of how the function is called as well as providing a guarded statement to allow different actions to be executed given certain system conditions. For instance, if an input is undefined when a function is run the guarded statement can enforce a different action. Without going into the details of the formal structure of RDSL™ we state that it provides a formal description of the state of the flows that an automated scheduler use to define a timed schedule with zero Heisenbugs by construction. This has dramatic implications on test time for RAN development.

Fig. 3.  RDSL™ Syntax for a Flow

In the example shown in Fig. 3 we show the definition of a flow which will, in turn, be called by other flows in this case. This flow has some internal streams defined, some of which are input and output streams and others purely internal. This flow calls other flows using an indexing syntax, which is a convenient way of producing multiple flows from a tensor stream. This is not an implementation but a declaration. How and when these streams are finally implemented is decided by the optimization process.

In the example in Fig. 4 a modifier is shown. Unlike a flow, a modifier is the bottom of the tree of calls and implements the call to a function. In this case a guarded structure is used to determine if there has been a timing error on the input so that a simple error message can be generated.

Fig. 4.  RDSL™ Syntax for a Modifier

### C.  Hardware Resources Abstraction

Hardware resources are defined in terms of the resource elements (i.e. processors, memories, interconnects, accelerators) and then a set of patterns are defined which describe how data buffers mover through the hardware from creation to destruction. An example of this is shown in Fig. 5, in xml format to allow us to collapse some of the fields. Each pattern has a name and is defined in terms of an anchor memory for definition and observation. This description was generated from a higher level topological map of the hardware, including the pattern names, which were generated for human readability. But once the hardware description has been generated, it can be hand edited to restrict pattern use to align it to the middleware used by the hardware to access the resources, for instance, in the case of Intel FlexRAN™, the eBBUPool middleware [19].

The other fields in the definition of this pattern describe its relationship to other patterns. Gabriel™ can reason about buffers that are observed multiple times by creating "sibling" buffers that can have their own pattern and then optimizing out redundant siblings at a later stage in the optimization.

Fig. 5.  Hardware Resource Syntax Example

The metadata for the individual functions (i.e. tasks) are similarly defined in terms of patterns as shown in Fig. 6. Each function is defined in terms of the patterns that can be used to manage the data it creates. Other metadata about runtime and memory use is also included and this can be a constant or an equation, referencing other values so that a very flexible map of runtime and memory use can be built up.

Fig. 6.  Hardware Resource Syntax Example

## IV. EXAMPLE RAN CONSTRUCTION AND PERFORMANCE RESULTS

The automation framework described in this paper has been applied to the Intel® FlexRAN™ solution [10] using Cirrus360's RAN automation platform Gabriel™ and RDSL™. The 5G NR protocol real time behavior and flows have been represented to Gabriel™ using RDSL™ as well as an abstract description of the Intel® Xeon Gold along with the Intel® RAN Accelerator ACC100 [15].

We used benchmarks from the FlexRAN™ test suite to create a hardware model for a particular test case within the suite. System constraints were applied to support the requirements of the test. In this way we can compare the performance of the FlexRAN™ platform solution, which has been carefully and expertly optimized by hand to meet the requirements of the complete test suite, with that of a deployment-optimized RAN constructed in an automated manner using the described DSL in this paper.

Several tests were conducted spanning sub-6 and mMIMO scenarios with a range of target deployment constraints such as latency. The automation platform analyzed and explored the solution space to find a feasible schedule of the software on the given hardware, such that the target deployment constraints were met, while the solution was optimized for maximum power savings opportunities. Using this methodology, the automation framework was able to significantly increase the power savings opportunity in the optimized FlexRAN™ solution compared to the manually optimized default configuration of FlexRAN™.

First, we present a test case of 4 cells downlink and uplink, 20MHz, supported on an Intel server platform with software Forward Error Correction. We modified the eBBUPool of the FlexRAN™ platform slightly to allow it to absorb a configuration file that is automatically generated by the Gabriel™ optimization platform, using the RDSL™ inputs, including different optimization constraints. In one case we ran Gabriel™ with a constraint to maximize the power-saving opportunity by minimizing the active period of the processing per slot.

The results are shown in Fig. 7 compared to the original FlexRAN™ solution. Note that the FlexRAN™ solution has been hand-optimized to complete each slot's computation in time for all of the test cases in the test suite. So, it does not have the same optimization criteria, and critically, it is not optimizing for this specific use case. Optimizing by hand each use case would be unfeasibly difficult, but with automation enabled by RDSL™ it becomes a realistic goal.

The results in Fig. 7 show that it is possible to reduce the activity period by 152us for this given deployment, an improvement in power savings of about 15%.

Our second test case is a massive MIMO 100MHz use case on the same platform but with the AC100 hardware accelerator for FEC. Fig. 8 is a visual representation of the timing of streams in the mMIMO example. The RDSL™ driven automation has decided on the timing and scheduling for this use case, a task that is today done manually by a team of experts.

The benefit can be seen in Fig. 9 where the runtime for the default FlexRAN™ solution is compared to the use case automated RDSL™ driven solution. There is a 26.7% improvement in uplink latency in this case.

Note that the performance benefit comes on top of the automation benefit. The Gabriel™ solution can automatically optimize in reaction to a change in the system or network requirements without intervention by a team of experts. This is especially important in cases where the RAN may fail a new use case and need to be hand re-optimized.

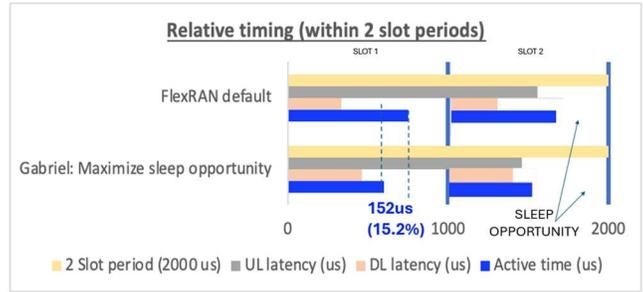

Fig. 7. Power saving comparison with deployment-specific optimization

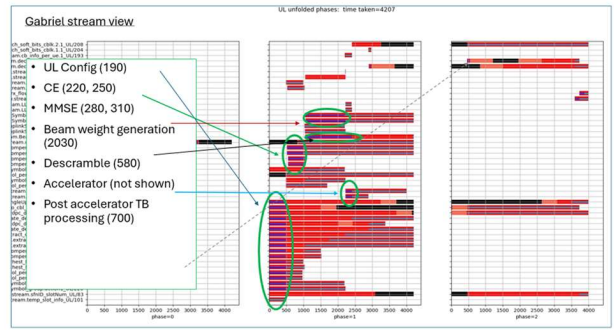

Fig. 8. Stream Timing for mMIMO Example

| KPI | Before & After | | Impact |
|---|---|---|---|
| Uplink Latency | FlexRAN default: | 1522us | 26.7% improvement in UL latency |
| | Use case optimized by Gabriel™ AUTOMATION: | 1115us | |

Fig. 9. Latency Comparison with Deployment Specific Optimization

As an example of how significant changes can be represented clearly within the declarative language of RDSL™ we show two code fragments in Fig. 10 and Fig. 11. The first fragment, Fig. 10, implements a flow which processes streams of data such that several channel compensation operations are declared independently. In the other fragment, Fig. 11, all of the channel compensation occurs within a single logical function. This algorithm difference is often seen in military applications when optimizing for high Doppler versus medium to low doppler, for instance non-terrestrial networks [13]. Apart from swapping out these two flow definitions, nothing else was changed in the RDSL™ description of uplink 5G mMIMO processing. Gabriel™ processed the RDSL™ declaration of the uplink functionality, along with constraints and the abstract description of the hardware. The conclusion, as we have seen in many examples implemented using FlexRAN™ reference software and RDSL™, is that a change in algorithm can be easily accommodated using the methodology described in this paper, even when a significant amount of rescheduling of functions is required in the FlexRAN™ reference software to meet latency or other constraints. This allows the FlexRAN™ reference software to be optimized for very specific deployment requirements, potentially saving power and cost in the network. For this example the two different flows can both be part of the RDSL™ declaration and the relative amount of each type can be added as a constraint, so balancing high and low doppler support for a specific deployment. Such fine granularity deployment tradeoffs are only possible with the automation approach described in this paper.

Fig. 10. RDSL™ for mMIMO in high Doppler case

Fig. 11. RDSL™ for mMIMO in low Doppler case

## V. Conclusion and Future Work

We have described a new language called RDSL™ that can be used to describe the requirements and data flow for the DU of an O-RAN. The same methodology could be applied to O-RU and O-CU as they all exhibit real time period properties. RDSL™ is a simple language that takes in constraints and requirements using a Kubernetes like syntax to allow it to integrate well with other DevOps tools. We have demonstrated the ability of an optimization platform for RDSL™ to produce a valid schedule with superior performance to a hand-optimized schedule as it can focus on the exact deployment requirements. Applying this language and automation methodology to CU and RU will allow for a holistic view of the complete RAN optimization. It will also allow for the addition of new features to an existing 5G RAN that are specific to a military communications application.

## Acknowledgment

We would like to thank Intel[1] and Vodafone for their unwavering support and advice in the development of our automated approach to RAN deployment, TIP for their support of the early development of RDSL™, and the NTIA and USDA for their continued support of our efforts.